\documentclass[10pt,twoside]{hsqcd}
\usepackage{epsfig}
\usepackage{epsf,amsmath}
\usepackage{graphicx}
 \newcommand{\be}{\begin{eqnarray}}
 \newcommand{\ee}{\end{eqnarray}}
 \newcommand{\eeq}{\end{equation}}
 \newcommand{\ba}{\begin{array}{1}}
 \newcommand{\ea}{\end{array}}
 \newcommand{\bb}{\begin{thebibliography}}
 \newcommand{\eb}{\end{thebibliography}}

\begin{document}

\title{FORWARD HEAVY BARYON PRODUCTION IN $pp$ COLLISIONS AT LHC }

\author{D.~A~.Artemenkov, V.~A. Bednyakov, \underline{G.~I.~Lykasov$^1$} \\ \\
JINR, Dubna, Moscow region, 141980,  Russia, \\
$^1$E-mail: lykasov@jinr.ru }

\maketitle

\begin{abstract}
\noindent We present the theoretical results on the forward
$\Lambda_b$ production in $pp$ collisions obtained within the soft QCD,
namely the quark gluon string model, at LHC energies. It 
can give us useful information on the Regge trajectories of 
the bottom mesons.
\end{abstract}



\markboth{\large \sl \underline{ G. Lykasov} \& D.Artemenkov, V. Bednyakov
\hspace*{2cm} HSQCD 2010} {\large \sl \hspace*{1cm} 
FORWARD HEAVY BARYON PRODUCTION IN $PP$ COLLISIONS ...} 


As is well known, there are successful phenomenological approaches for describing 
the soft hadron-nucleon,
hadron-nucleus and nucleus-nucleus interactions at high energies 
based on the Regge theory and the $1/N$ 
expansion in QCD, for example the quark-gluon string model (QGSM) \cite{kaid1}.
and the dual parton model (DPM) \cite{Capella:1994}. 
In this paper we present the results on the beauty baryon production, in particular 
$\Lambda_b$, in $pp$ collisions at LHC energies and small $p_t$ within the QGSM to find
the information on the Regge trajectories of the bottom ($b{\bar b}$) mesons and 
the fragmentation functions (FF) of all the quarks and diquarks to this baryon. 
Actually, these results are the predictions for the LHC experiments.
The general form for the invariant inclusive hadron spectrum
within the QGSM is \cite{kaid1,Capella:1994}
\begin{eqnarray}
E\frac{d\sigma}{d^3{\bf p}}\equiv
\frac{2E^*}{\pi\sqrt{s}}\frac{d\sigma}{d x d p_t^2}=
\sum_{n=1}^\infty \sigma_n(s)\phi_n(x,p_t)~, 
\label{def:invsp}
\end{eqnarray}
where $E,{\bf p}$ are the energy and the three-momentum of the
produced hadron $h$ in the laboratory system (l.s.); 
$E^*,s$ are the energy of $h$ and the square of the initial energy in the
c.m.s of $pp$; $x,p_t$ are the Feynman variable and the transverse
momentum of $h$; $\sigma_n$ is the cross section for production of
the $n$-Pomeron chain (or $2n$ quark-antiquark strings) decaying
into hadrons, calculated within the quasi-``eikonal approximation''
\cite{Ter-Mart}. Actually, the function $\phi_n(x,p_t)$ is the convolution
of the quark (diquark) distributions and the FF, see the details 
in \cite{kaid1} and \cite{Capella:1994,kaid2}.
\be
\phi^{p p}_n(x)=F_{qq}^{(n)}(x_+)F_{q_v}^{(n)}(x_{-})+
F_{q_v}^{(n)}(x_+)F_{qq}^{(n)}(x_-)+\\
\nonumber
2(n-1)F_{q_s}^{(n)}(x_+)F_{{\bar q}_s}^{(n)}(x_-)~,
\ee
where
$x_{\pm}=\frac{1}{2}(\sqrt{x^2+x_t^2}\pm x)$~, 
and
\be
F_\tau^{(n)}(x_\pm)=\int_{x_\pm}^1 dx_1 f_\tau^{(n)}(x_1)G_{\tau\rightarrow h}
\left(\frac{x_\pm}{x_1}\right)~,
\label{def:Ftaux}
\ee
 Here $\tau$ means the flavour of the valence (or sea) quark or diquark, $f_\tau^{(n)}(x_1)$
is the quark distribution function depending on the longitudinal momentum fraction $x_1$  
 in the $n$-Pomeron chain; $G_{\tau\rightarrow h}(z)=
z D_{\tau\rightarrow h}(z)$, $ D_{\tau\rightarrow h}(z)$ is the FF of a quark (antiquark) or 
diquark of flavour $\tau$ into a hadron $h$ (charmed or bottom hadron in our case).

All the details of the calculation of Eq.(\ref{def:invsp}) and the
interaction function $\phi_n(x,p_t)$ can be found in \cite{LLB:2010,BLL:2010}.
 Fig.1 illustrates the sensitivity of the inclusive
spectrum $d\sigma/dx$ of the produced charmed  baryons $\Lambda_c$ to the different values for 
the intercept $\alpha_\psi(0)$ of the $\Psi(c{\bar c})$ Regge trajectory. 
The solid line corresponds to $\alpha_\psi(0)=0$, and the dashed curve corresponds to 
$\alpha_\psi(0)=-2.18$. One can see that there are difference between the R608 \cite{R608} and
R422\cite{R422} experimental data. As is shown in \cite{BLL:2010},  the R608 data are more 
adequate to our calculations at $\alpha_\Psi(0)=0$, see the solid line in Fig.1.
The $\Lambda_b$ baryon produced in $pp$ collision 
can decay $\Lambda_b\rightarrow J/\Psi \Lambda^0$
 with the branching ratio $Br_{\Lambda_b\rightarrow J/\Psi\Lambda^0}=
\Gamma_{\Lambda_b\rightarrow J/\Psi\Lambda^0}/\Gamma_{tot}=(4.7\pm 2.8)\cdot 10^{-4}$ 
and $J/\Psi$ decays into $\mu^+\mu^-$ ($Br_{J/\Psi\rightarrow\mu^+\mu_-}=(5.93\pm 0.06)\%$)
 or into $e^+e^-$ ($Br_{J/\Psi\rightarrow e^+e^-}=5.93\pm 0.06\%$), whereas $\Lambda^0$ can decay into 
$p\pi^-$ ($Br_{\Lambda^0\rightarrow p\pi^-}=63.9\pm 05\%$), or into $n\pi^0$ 
($Br_{\Lambda^0\rightarrow n\pi^0}=35.8\pm 0.5\%$).
Experimentally one can measure the differential cross section
$d\sigma/d\xi_p dt_p dM_{J/\Psi}$, where $\xi_p=\Delta p/p$ is the energy loss, $t_p=(p_{in}-p_1)^2$ is 
the four-momentum transfer, $M_{J/\Psi}$ is the effective mass of the $J/\Psi$-meson. 
\begin{figure}[ht]
   \begin{center}
 {\epsfig{file=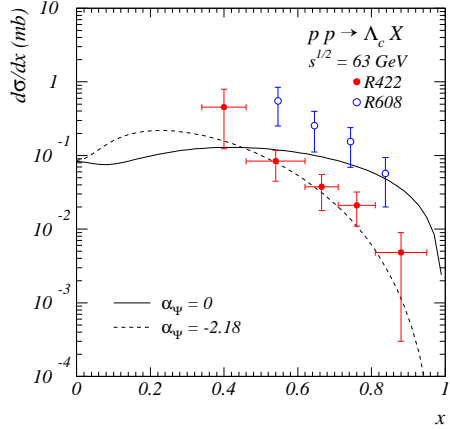,width=0.45\linewidth  }}
\caption[Fig.3]{The differential cross section $d\sigma/dx$ for the
 inclusive process $pp\rightarrow\Lambda_c X$ at $\sqrt{s}=62~\mathrm{GeV}$.
 The solid line corresponds to $\alpha_\psi(0)=0$, and the dashed curve corresponds to 
$\alpha_\psi(0)=-2.18$. The open circles correspond to the R608 experiment \cite{R608},
 and the dark circles correspond to the R422 experiment \cite{R422}.} 
\end{center}
 \end{figure}
\label{sec:figures}
\begin{figure}[hb]
\centerline{\includegraphics[width=0.45\textwidth]{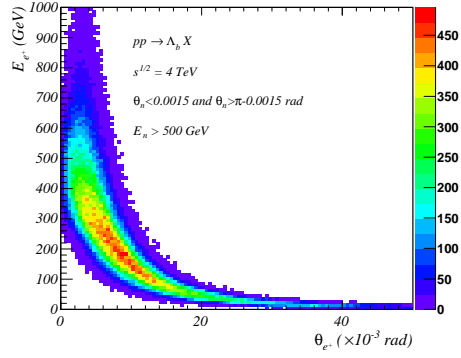}}
\caption{The distribution over $\theta_{e^+}$ and $E_{e^+}$ in the 
inclusive process $pp\rightarrow\Lambda_b X\rightarrow J/\Psi\Lambda^0 X
\rightarrow e^+e^- n\pi^0 X$ at $\sqrt{s}=$4 GeV. The fraction of the 
events is about 4.6 percent (13.8 nb).}\label{Fig:MV}
\end{figure}
\label{sec:figures}
\begin{figure}[hb]
\centerline{\includegraphics[width=0.8\textwidth]{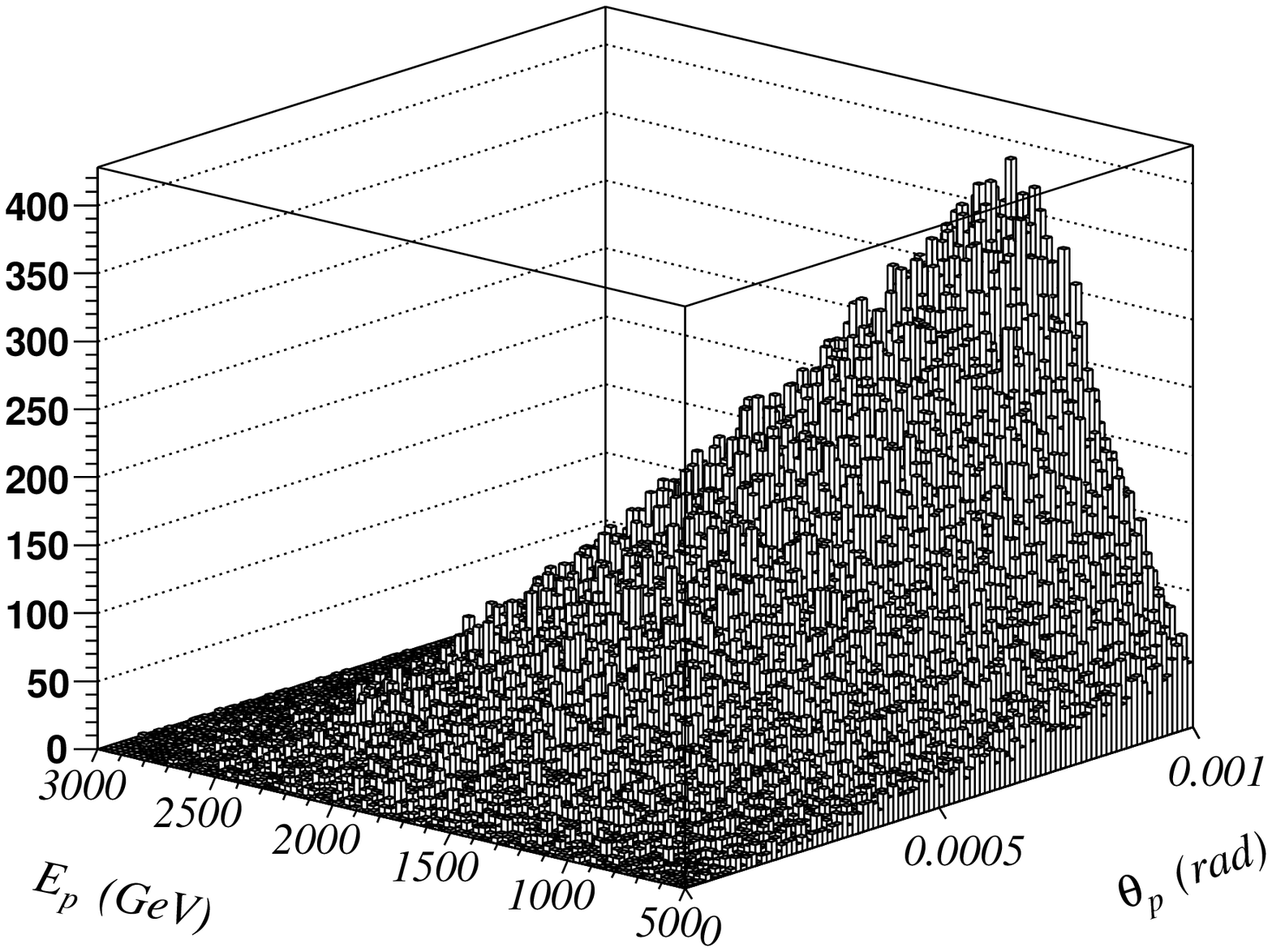}}
\caption{The two-dimensional distribution 
 over $\theta_p$ and $E_p$ in the
 inclusive process $pp\rightarrow\Lambda_b X\rightarrow J/\Psi\Lambda^o X\rightarrow e^+e^- p\pi^- X$ 
at $\sqrt{s}=10~\mathrm{TeV}$ at $\alpha_\Upsilon(0)=0$, when $E_p\ge 500 GeV$ and $\theta_p\leq 1 mrad.$. 
The rate of these events is about 0.74 percent (2.22 nb).}
\label{Fig:MV}
\end{figure}

\label{sec:figures}
\begin{figure}[hb]
\centerline{\includegraphics[width=1.0\textwidth]{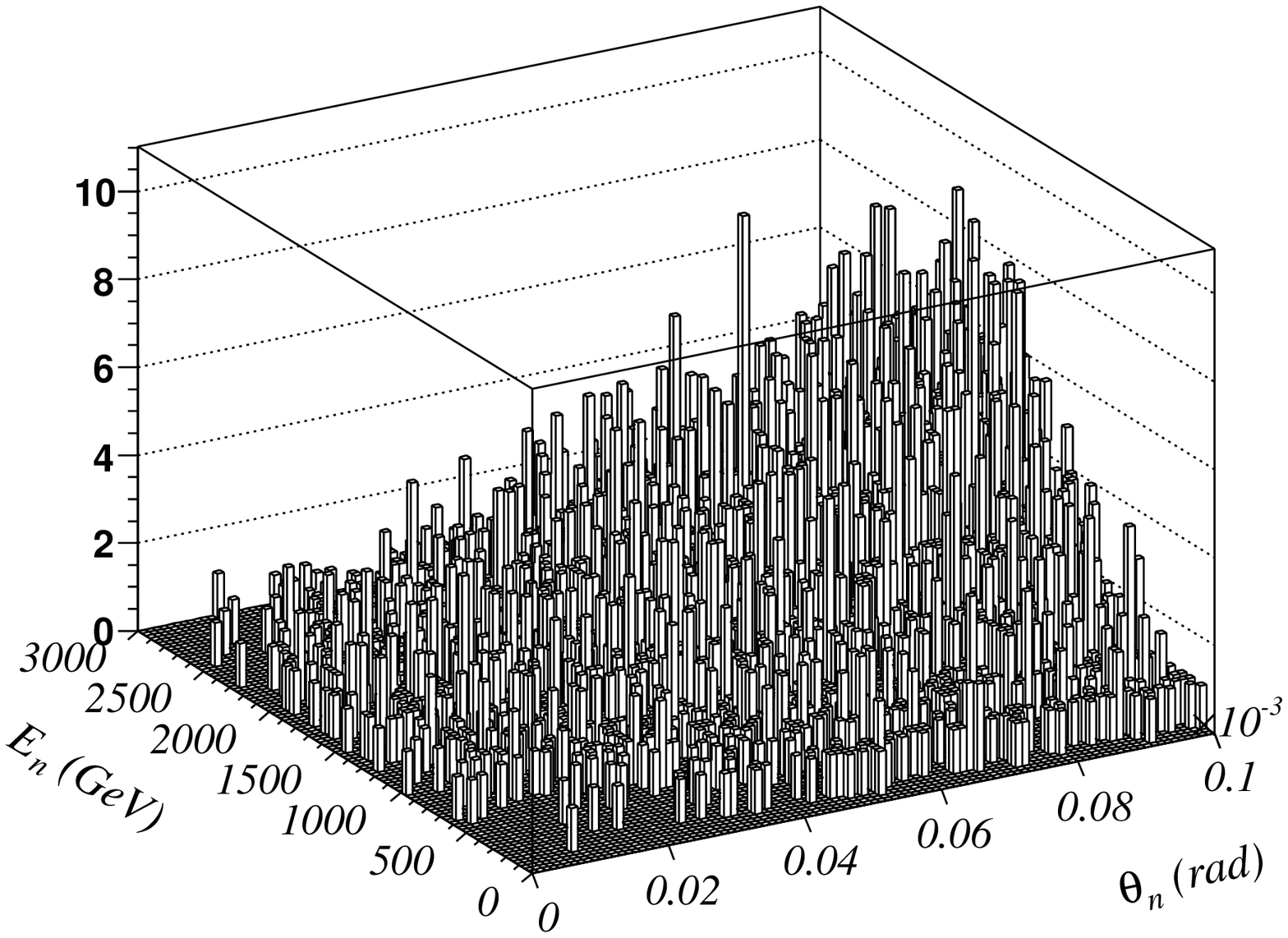}}
\caption{The two-dimensional distribution over $\theta_p$ and $E_p$ in the
 inclusive process $pp\rightarrow\Lambda_b X\rightarrow J/\Psi\Lambda^o X\rightarrow e^+e^- n\pi^0 X$ 
at $\sqrt{s}=10~\mathrm{TeV}$ at $\alpha_\Upsilon(0)=0$, when $\theta_n\leq 0.1 mrad.$.The rate of these events 
is about 0.015 percent (45 pb).}
\label{Fig:MV}
\end{figure}
The detailed predictions on these reactions are presented in \cite{BLL:2010}, where it is shown
that all the observables are very sensitive to the value of intercept $\alpha_\Upsilon(0)$ 
of the $\Upsilon(b{\bar b})$ Regge trajectory. The upper limit of our results is reached at  $\alpha_\Upsilon(0)=0$,
when this Regge trajectory as a function of the transfer $t$ is nonlinear.
 Using the hadron detector at the CMS and the TOTEM one could register the decay
$\Lambda^0_b\rightarrow J/\Psi~\Lambda^0\rightarrow \mu^+\mu^-~\pi^- p$
by detecting two muons and one proton emitted forward. 
However, the acceptance of the muon detector
is $10 ^0\leq\theta_\mu\leq 170^0$ \cite{ATLAS1}, where, according to our calculations,
the fraction of this events is to low. On the other hand, the electromagnetic calorimeter
at the CMS is able to measure the dielectron pairs $e^+e^-$ in the acceptance about
$1^0\leq\theta_{e(e^+)}\leq 179^0$ \cite{CMS}.
Fig.2 illustrates that the electrons and positrons produced from the $J/\Psi$ decay are 
emitted at very small scattering angles, mainly at $\theta_e<16 mard$. The rate of these events, 
when the neutrons are emitted at $\theta_n<1.5 mrad$ and $E_n>500$ GeV, is about 4.6 percent (13.8 nb). 
In Fig.3 the two-dimensional distribution over $E_p$ and $\theta_p$ for the reaction
$pp\rightarrow\Lambda_b X\rightarrow J/\Psi\Lambda^o X\rightarrow e^+e^- p\pi^- X$ is presented. 
The rate of these events is about 0.74 percent (2.22 nb).
This could 
be reliable using the TOTEM together with the CMS \cite{Deile}. 

The ATLAS is able also to detect $e^+e^-$  by the electromagnetic calorimeter in the
interval $1^0\leq\theta_{e(e^+)}\leq 179^0$ \cite{ATLAS1} and the neutrons emitted forward at the
angles $\theta_n\leq 0.1 mrad$ \cite{ZDC}.
In Fig.4 we present the prediction for the reaction 
$pp\rightarrow\Lambda_b X\rightarrow J/\Psi\Lambda^o X\rightarrow e^+e^- n\pi^0 X$,
that could be reliable at the ATLAS experiment.
The rate of these events is about 0.015 percent (45 pb).

The TOTEM~\cite{TOTEM} together with the CMS might be able to measure the channel
$\Lambda_b\rightarrow J/\Psi~\Lambda^0\rightarrow e^+e^-~\pi^- p$
(the integrated cross-section is about 0.2-0.3 $\mu$b at $\alpha_\Upsilon(0)=0$ and
smaller at $\alpha_\Upsilon(0)=-8$).
The T2 and T1 tracking stations of the TOTEM apparatus have their angular acceptance 
in the intervals $\rm 3\,mrad < \theta < 10\,mrad$ (corresponding to $6.5 > \eta > 5.3$) 
and $\rm 18\,mrad < \theta < 90\,mrad$ (corresponding to $4.7 > \eta > 3.1$) 
respectively, 
and could thus detect 42\% of the muons from the $J/\Psi$ decay.
In the same angular intervals, 36\% of the $\pi^{-}$ and 35\% of 
the protons from the $\Lambda^0$ decay are expected. 
According to a very preliminary estimate~\cite{Deile}, protons with energies 
above 3.4\,TeV
emitted at angles smaller than 0.6\,mrad 
could be detected 
in the Roman Pot station at 147\,m from IP5 \cite{TOTEM,Deile}. In the latter case, 
the reconstruction of the proton kinematics may be possible, whereas 
the trackers T1 and T2 do not provide any momentum or energy information.
Future detailed studies are to establish the full event topologies with
all correlations between the observables in order to assess whether the 
signal events can be identified and separated from backgrounds. These 
investigations should also include the CMS calorimeters HF and CASTOR which 
cover the same angular ranges as T1 and T2 respectively \cite{Deile}.

 We analyzed the production of charmed and beauty baryons in proton-proton collisions at high energies 
within soft QCD, namely the quark-gluon string model (QGSM). This approach can describe 
rather satisfactorily the charmed baryon production in $pp$ collisions \cite{LLB:2010,BLL:2010}. 
It allows us to apply the QGSM to studying the beauty baryon production in $pp$ collisions. 
We focus mainly on the analysis of the forward $\Lambda_b$ 
production in $pp$ collisions at LHC energies and got some predictions which could be reliable at the
TOTEM and ATLAS experiments at CERN.  
 We present the predictions for the reaction
$pp\rightarrow\Lambda_b X\rightarrow e^+e^- p\pi^- X$ 
that could be reliable at the TOTEM together with the CMS, and for the process
$pp\rightarrow\Lambda_b X\rightarrow e^+e^- n\pi^0 X$
which can be reliable at the ATLAS experiment using the ZDC. 
We did not include the diffractive and double diffractive $\Lambda_b$ production
in $pp$ collisions because these processes can be experimentally
separated from the forward $\Lambda_b$ production
\cite{TOTEM,Deile}. 

Note  that in this paper we neglect the contribution of the intrinsic charm in the proton calculating the 
charmed baryon production in $pp$ collisions  
and possible intrinsic beauty in the proton. 
However, as shown 
recently \cite{Ullrich:2010}, the intrinsic charm in the proton can result in a sizable contribution to the forward 
charmed meson production. 
As is shown in \cite{Pumplin:2006}, the probability to find the intrinsic charm in the proton is not more than
0.5 percent. However, the probability of the intrinsic bottom in the proton is suppressed by a factor
 $m^2_c/m^2_b\simeq 0.1$
\cite{Polyakov:1999}, where $m_c$ and $m_b$ are the masses of the charmed and bottom quarks. Therefore,
the contribution of the intrinsic bottom in the proton to the discussed reaction can be suppressed in comparison
to the intrinsic 
charm contribution about ten times. Anyway, it would be interesting to study this problem more carefully.

\vspace{0.5cm}
 We are very grateful to V.V.Lyubushkin for a help in the MC calculations.
We also thank  M. Deile, P. Grafstr{\"o}m, and  N.I. Zimin    
for extremely useful help related to the possible experimental check of the
suggested predictions at the LHC and the preparation of this paper.
We are also grateful to  D.Denegri, K. Eggert,\frame{A. B. Kaidalov}, O.V.Teryaev, 
 M. Poghosyan and S.White for very useful discussions. 
This work was supported in part by the Russian Foundation for Basic Research 
grant N: 08-02-01003.



\begin{footnotesize}

\end{footnotesize}



\begin{thebibliography}{99}


\bibitem{kaid1}
  A.~B.~Kaidalov and K.~A.~Ter-Martirosyan,
  Phys. Lett. B {\bf 116} (1982) 489,
  [arXiv:hep-ph/0909.5061].
\bibitem{Capella:1994}
  A.~Capella, U.~Sukhatme, C.~I.~Tan and J.~Tran Than Van  
  Phys. Rep. {\bf 236} (1994) 225,
  [arXiv:hep-ph/0909.5061].
\bibitem{Ter-Mart}
K.~A.~Ter-Martirosyan,
Phys. Lett. B{\bf 44} (1973) 377. 
\bibitem{kaid2}
A.~B.~Kaidalov \and O.~I.~Piskunova, 
Z.Phys. C{\bf 30}(1986)145.
\bibitem{LLB:2010}
  G.~I.~Lykasov, V.~V.~Lyubushkin and V.~A.~Bednyakov,
  Nucl. Phys. [Proc. Suppl.]{\bf 198} (2010) 165
  [arXiv:hep-ph/0909.5061].
\bibitem{BLL:2010}
  V.~A.~Bednyakov, G.~I.~Lykasov and V.~V.~Lyubushkin,
  arXiv:hep-ph/1005.0559.
\bibitem{ATLAS1}
ATLAS Collaboration, 
Technical Design Report,
ATLAS-TDR-017,CERN-LHCC-2005-022.
\bibitem{CMS}
CMS Collaboration 
J. Phys. G: Nucl. Part. Phys. {\bf 34} (2007) 995. 
\bibitem{R608}
P.~Chauvat, et al. Phys.Lett. B{\bf 199} (1987) 304.
\bibitem{R422}
G.~Bari, et al. Nuovo Cim. A{\bf 104}(1991) 571.
\bibitem{Deile} 
M.~Deile, 
Private communication; 
H.~Niewiadomski,
TOTEM-NOTE, {\bf 002} (2009).
\bibitem{ZDC}
ATLAS Collaboration,
Letter of Intent ``Zero Degree Calorimeters''. 
\bibitem{TOTEM} 
TOTEM Collaboration,
Technical Design Report, (2004),CERN-LHCC-2004-002; 
Addendum CERN-LHCC-2004-020, 
``The Totem Experiment At The CERN Large Hadron Collider'', JINST {\bf 3}{2008} 
S08007.
\bibitem{Ullrich:2010}
V.~P.~Goncalves, F.~S.~Navarra and T.~Ullrich, 
arXiv:hep-ph/0805.0810.
\bibitem{Pumplin:2006}
J.~Pumplin,
Phys.Rev. D{\bf 73} (2006) 114015.
\bibitem{Polyakov:1999}
M.~V.~Polyakov, A.~Shafer, O.~V.~Teryaev,
Phys.Rev. D{\bf 60} (1999) 051502.
\end{thebibliography}
\end{document}